\documentclass[aps,prb,twocolumn,superscriptaddress,longbibliography]{revtex4-1}
\usepackage{graphicx}
\usepackage{dcolumn}
\usepackage{float}
\usepackage{color}
\usepackage[textwidth=2cm, disable]{todonotes}
\begin{document}

\title{Re-entrant superspin glass phase in La$_{0.82}$Ca$_{0.18}$MnO$_3$ ferromagnetic insulator }

\author{P. Anil Kumar}
\affiliation{Centre for Advanced Materials, Indian Association for the Cultivation of Science, Kolkata 700032, India}
\affiliation{Department of Engineering Sciences, Uppsala University, P.O. Box 534, SE-751 21 Uppsala, Sweden}
\author{R. Mathieu}
\affiliation{Department of Engineering Sciences, Uppsala University, P.O. Box 534, SE-751 21 Uppsala, Sweden}
\author{P. Nordblad}
\affiliation{Department of Engineering Sciences, Uppsala University, P.O. Box 534, SE-751 21 Uppsala, Sweden}
\author{Sugata Ray}
\affiliation{Centre for Advanced Materials, Indian Association for the Cultivation of Science, Kolkata 700032, India}
\affiliation{Department of Materials Science,Indian Association for the Cultivation of Science, Kolkata 700032, India}
\author{Olof Karis}
\affiliation{Department of Physics and Astronomy, Uppsala University, Box - 516, 75120 Uppsala, Sweden}
\author{Gabriella Andersson}
\affiliation{Department of Physics and Astronomy, Uppsala University, Box - 516, 75120 Uppsala, Sweden}
\author{D. D. Sarma}
\email[sarma@sscu.iisc.ernet.in]{}
\affiliation{Centre for Advanced Materials, Indian Association for the Cultivation of Science, Kolkata 700032, India}
\affiliation{Department of Physics and Astronomy, Uppsala University, Box - 516, 75120 Uppsala, Sweden}
\affiliation{Solid State and Structural Chemistry Unit, Indian Institute of Science, Bangalore 560012, India}
\affiliation{Also at Jawaharlal Nehru Centre for Advanced Scientific Research, Bangalore and Council of Scientific and Industrial Research-Network of Institutes for Solar Energy (CSIR-NISE).}

\date{\today}

\begin{abstract}
We report results of magnetization and ac susceptibility measurements down to very low fields on a single crystal of the perovskite manganite, La$_{0.82}$Ca$_{0.18}$MnO$_3$. This composition falls in the intriguing ferromagnetic insulator region of the manganite phase diagram. In contrast to earlier beliefs, our investigations reveal that the system is magnetically (and in every other sense) single-phase with a ferromagnetic ordering temperature of $\sim$ 170 K. However, this ferromagnetic state is magnetically frustrated, and the system exhibits pronounced glassy dynamics below 90 K. Based on measured dynamical properties, we propose that this quasi-long-ranged ferromagnetic phase, and associated superspin glass behavior, is the true magnetic state of the system, rather than being a macroscopic mixture of ferromagnetic and antiferromagnetic phases as often suggested. Our results provide an understanding of the quantum phase transition from an antiferromagnetic insulator to a ferromagnetic metal via this ferromagnetic insulating state as a function of $x$ in La$_{1-x}$Ca$_x$MnO$_3$, in terms of the possible formation of magnetic polarons.
\end{abstract}


\maketitle

\section{Introduction}
Hole-doped perovskite manganites of general formula $A_{1-x}B_x$MnO$_3$ ($A$ = trivalent Lanthanide, $B$ = divalent alkali metal) with a low doping, generally 0.05 $< x <$ 0.22 for $A$ = La and $B$ = Ca, are of fundamental interest because they constitute the few examples of ferromagnetic insulators, unlike the ones with higher $x$ or larger bandwidth which are ferromagnetic but also metallic, or with a lower $x$ which are insulating, but also antiferromagnetic.\cite{Dagotto20011} While the origin of ferromagnetic, insulating state in some undoped compounds, such as La$_{2}$NiMnO$_6$, \cite{Das2008186402,Choudhury2012127201} has been explained in the past, charge doped systems present additional difficulties in comprehending a ferromagnetic insulating ground state. Specifically, the origin of the coexistence of ferromagnetic and insulating properties in La$_{1-x}$Ca$_x$MnO$_3$  (LCMO), has not been established. The ferromagnetic metal (FMM) phase of the perovskite manganites with larger doping has been successfully  explained by considering the double exchange mechanism proposed by Zener. \cite{Zener1951403,deGennes1960141} This mechanism, while explaining the ferromagnetism, invariably requires the system to be metallic due to hopping of electrons between Mn$^{3+}$ and Mn$^{4+}$ and hence failing to explain the ferromagnetic insulating state. Further it is to be noted that electronic phase separation of varying length scales has been reported at higher Ca doping levels.\cite{Shenoy20062053} \par A major obstacle in establishing the true ground state in case of manganites has been the difficulty in preparing single phase samples, which has lead to the speculation \cite{Markovich2002144402,Markovich2002094409} that the ferromagnetic insulating (FMI) state is the result of spatially distinct coexistence of separate ferromagnetic metallic and antiferromagnetic insulating phases in a single sample. However, investigations performed using other single crystal samples have suggested microscopically homogeneous electronic properties of the samples.\cite{Dai2001224429,Dai20002553,Jiang2009214433} Interestingly, even the reports on these homogeneous samples are interpreted to promote contrasting pictures for the magnetic ground state. For example, the authors of Refs.[\citenum{Dai2001224429,Dai20002553}] observed a non-diverging magnetic correlation length and signatures of short range magnetic polarons using neutron scattering experiments, while a much more recent work\cite{Jiang2009214433} suggested an ideal 3-dimensional Heisenberg ferromagnetic ground state, based on the values of critical exponents that they obtain for the FMI composition of LCMO. Nevertheless, both theory and experiments suggest formation of local lattice distortions or lattice polarons in the ordered state leading to a nanoscale inhomogeneity which is starkly different from the chemical or macroscopic electronic phase separation. \cite{Dai20002553,Mathieu2004227202,Sarma2004097202,Mathieu2007124706,Cengiz2007127202,Shenoy2007097201} \par AC susceptibility technique can be useful in determining the true magnetic state of a material. An enlightening report on the magnetic properties of a single crystal of La$_{0.8}$Ca$_{0.2}$MnO$_3$, which is closer to the FMM composition, using ac-techniques is found in Ref. [\citenum{Hong20021583}] however without considering non-equilibrium (aging) and non-linear field effects. We here investigate in detail the static and dynamical magnetic properties of a single-crystal of La$_{0.82}$Ca$_{0.18}$MnO$_3$ (LCMO18) whose composition lies within the FMI regime (0.05 $< x <$ 0.22). It is found that the material orders ferromagnetically, albeit short ranged and displaying glassy dynamics. The observed glassy dynamics suggest that frustration effects govern the ferromagnetic configuration, as in re-entrant spin glass systems (re-entrant ferromagnet in the present case). However, with the significant difference that the magnetic entities in the present case are groups of spins (magnetic polarons), instead of the single spins of conventional re-entrant magnets or spin glasses; i.e. a re-entrant superspin glass state. The results indicate that the re-entrant ferromagnetic insulating state is the intrinsic magnetic state of this composition, rather than a (macroscopically) phase separated one. We discuss the microscopic nature and origin of the ferromagnetic insulating phase, as well as the transition to ferromagnetic metal for larger hole doping.

\section{Experimental Details}
La$_{0.82}$Ca$_{0.18}$MnO$_3$ single crystal pieces were grown by floating zone technique (Crystal Systems Corporation, Japan) starting from a single phase polycrystalline sample of the same composition. To compensate for the Mn evaporation during crystal growth 1\% MnO is additionally added. The phase purity of the final grown crystal is verified by powder x-ray diffraction (XRD) using a Bruker D8 advance diffractometer. Spot analysis of energy dispersive x-ray spectroscopy on a Jeol FESEM reveals that the cation distribution in the plane perpendicular to the growth direction is uniform, and that the cation concentrations are also close to the nominal composition. The correct stoichiometry of the single crystals was additionally verified by Inductively Coupled Plasma-optical emission spectroscopy (ICP-OES) using a Perkin Elmer instrument. The electronic transport measurements are carried out using a laboratory setup and PPMS from Quantum Design Inc. PPMS is also used for heat capacity measurements. Magnetization $M$ and ac susceptibility (in-phase component $\chi{'}$ and out-of-phase component $\chi{''}$) data are collected, on a roughly rectangular piece of the crystal with 5 mm $\times$ 1.2 mm $\times$ 1.5 mm size, using Quantum Design MPMS XL SQUID magnetometer. Magneto-optic Kerr effect (MOKE) images were recorded at temperatures between 80 K and 273 K using a polarizing microscope with an external field applied in the plane of the sample, and parallel to the scattering plane of the light (longitudinal MOKE configuration).

\section{Results and Discussion}

\begin{figure}
\includegraphics{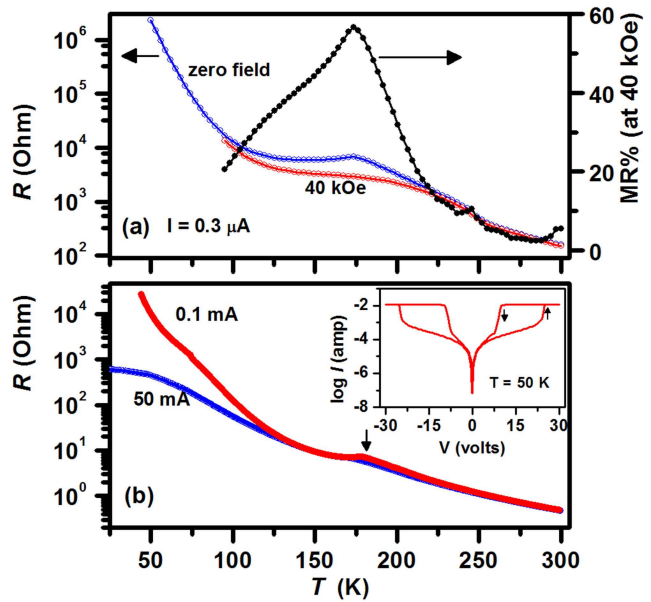}
\caption{\label{Fig.1 } (Color) (a) Temperature dependence of the electrical resistance $R$ of LCMO18 recorded under zero and 40 kOe fields. The third curve (right axis) shows the associated magnetoresistance $MR$\% = 100 $\times$ ($R$($H$ = 0)-$R$($H$ = 40 kOe))/$R$($H$ = 0) (b) The behavior of $R$ with temperature, $T$ for different excitation currents is shown to exemplify the electroresistance behavior. The inset in panel (b) shows a typical $I$-$V$ curve recorded at $T$ = 50 K. The flat region in the curve is due to the current limit, 10 mA, of the electrometer.}
\end{figure}
In Figure 1(a), we present the temperature dependence of the electrical resistance, $R$, of the LCMO18 single crystal.  A weak signature of insulator to metal transition can be observed in the zero magnetic field data at about 170 K before re-entering an insulating state on further lowering the temperature below $\sim$ 140 K. It is in agreement with the earlier literature data\cite{Wu20101705} for this composition, this fact also ensures the quality of the sample, since the electronic transport behavior is the most sensitive property to the oxygen non-stoichiometry and compositional variation. The associated magnetoresistance (\textit{MR}) observed in 40 kOe applied field is also indicated in the figure. Interestingly, this composition also exhibits colossal electroresistance (CER) \cite{Yuzhelevski2001224428,Jain2006152116} as demonstrated in Figure 1(b) and the inset. The resistance of the crystal is indeed highly sensitive to the magnitude of current that passes through the sample. The inset shows the reversible switching of sample resistance with applied voltage. Corresponding results on MR and CER of LCMO have been reported in the literature, however the details of the large electroresistance are different. \cite{Yuzhelevski2001224428,Jain2006152116}

\begin{figure}
\includegraphics{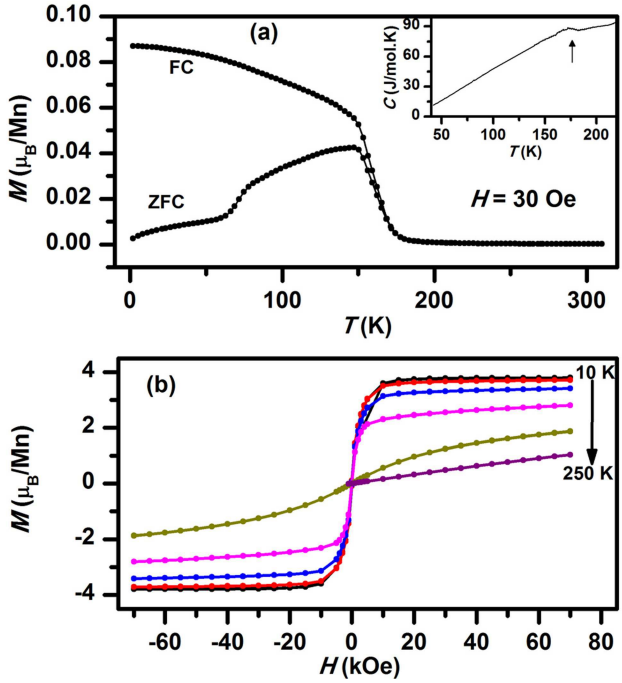}
\caption{\label{Fig.2 } (Color) (a) Temperature dependence of the zero-field-cooled (ZFC) and field-cooled (FC) magnetization in 30 Oe field. Inset shows the temperature dependence of the heat capacity $C$ measured on heating. (b) Magnetic field dependence of the magnetization (hysteresis loops) measured at different temperatures (10, 50, 100, 150, 200 and 250 K).}
\end{figure}
In Figure 2(a) we show the zero-field-cooled (ZFC) and field-cooled (FC) magnetization as a function of temperature. In the ZFC curve, a sharp rise in $M$ resembling a magnetic transition is observed near 170 K, coinciding with the insulator-metal transition temperature (c.f. fig. 1(a)). In addition, a drop in ZFC magnetization is observed at $\sim$ 70 K on lowering the temperature. This second anomaly is not reflected in the heat capacity ($C$) curve depicted in the inset to Figure 2(a). However a peak is observed in the $C$($T$) curve in the vicinity of the first magnetic transition, confirming the enhanced magnetic correlations in the system. The peak in heat capacity is relatively broad, suggesting that the long-range ordering of the ferromagnetic state established at 170 K is hindered.\cite{egilmez2008132505} Interestingly the magnetization - field (hysteresis) loops measured at different temperatures across the two anomalies in $M$($T$) curve, presented in Figure 2(b), show ferromagnetic behavior at and below 150 K in line with the magnetic ordering at 170 K. The saturation moment per Mn ($\sim$3.78 $\mu_B$) at 10 K is very close to the value (3.82 $\mu_B$) expected for a full ferromagnetic arrangement of Mn ions in the system. The temperature and magnetic field dependence of the magnetization hence suggests that the system is ferromagnetically ordered below 170 K. However, with hindered critical divergence at $T_C$. Yet, the high field magnetic moment (for fields larger than 10 kOe) corresponds to full polarization of that ferromagnetic state, suggesting the lack of macroscopic phase separation into ferromagnetic and antiferromagnetic regions. Further, Kerr microscopy images collected on our crystal at 100 K did not show any domain structure within the instrumental resolution limit of 0.5 $\mu$m. However, the magneto-optic-Kerr-effect (MOKE) intensity follows the bulk magnetization value as shown in Figure 3. The MOKE data corresponds to a 425 $\times$ 325 $\mu$m$^2$ area of the crystal with the field applied in the same plane as for the bulk magnetization measurement.

\begin{figure}
\includegraphics{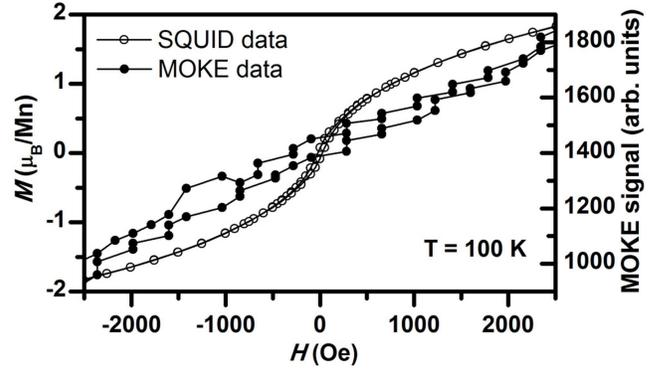}
\caption{\label{Fig.3 } Comparison of MOKE intensity to the bulk magnetization curve measured at 100 K.}
\end{figure}

\begin{figure}
\includegraphics{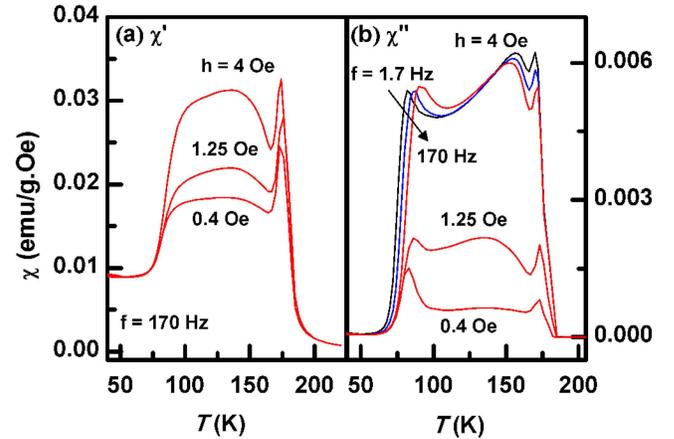}
\caption{\label{Fig.4 } (Color) Temperature dependence of the (a) in-phase $\chi{'}$ and (b) out-of-phase $\chi{''}$ components of the ac-susceptibility recorded using different amplitudes of the probing ac field ($h$ = 0.4, 1.25 and 4 Oe). The frequency $f$ is 170 Hz. $\chi{''}$(T) data obtained for 1.7 and 17 Hz are also plotted for $h$  = 4 Oe as indication of dynamical behavior.}
\end{figure}
The result of temperature dependent ac-susceptibility measurements is presented in Figure 4. Data collected using three different ac-field amplitudes, $h$, are shown.  As seen in the figure, the temperature dependence of the in- and out-of-phase components of the susceptibility is qualitatively similar for all amplitudes of the ac probing field. For example the $\chi{'}$($T$) curves include a sharp peak near 170 K, and a drop in the susceptibility below 90 K, akin to the features observed in the ZFC magnetization presented in Figure 2(a). Sharp peaks near 170 K and drops below 90 K are also observed for all amplitudes in $\chi{''}$($T$) curves. The low field susceptibility is strongly non-linear and enhanced by the amplitude of the ac field, $h$ (see e.g. the 100 - 150 K temperature interval for different values of $h$). Additional measurements as a function of amplitude $h$ show that the response of the system is essentially linear at least up to 0.5 Oe at temperatures below 140 K. We hence choose an amplitude of 0.4 Oe for all subsequent ac magnetic measurements in order to study the intrinsic magnetism of the material. For this amplitude, the ac-susceptibility curves (see left panels of Fig. 5) are reminiscent to those of re-entrant ferromagnets, which exhibit two peaks in the $\chi{''}$($T$) curves reflecting ferromagnetic and spin-glass phase transitions respectively, \cite{Jonason19966507,Mathieu2000441,Choudhury2012127201} the higher temperature transition with no frequency dispersion while the low temperature transition showing clear frequency dispersion. Spin glass states result from frustration effects due to the presence of competing magnetic interactions (the frustration may also be geometric). Further, particulates of ferromagnetic domains may also exhibit glassy magnetic features when the particulate size is low enough.\cite{Wu2003174408}  
It is also important to note that the spin correlation length does not diverge at the ferromagnetic transition temperature in the FMI composition of LCMO, suggesting formation of nanoscopic ferromagnetic domains at the transition temperature.\cite{Dai2001224429}

\begin{figure*}
\includegraphics{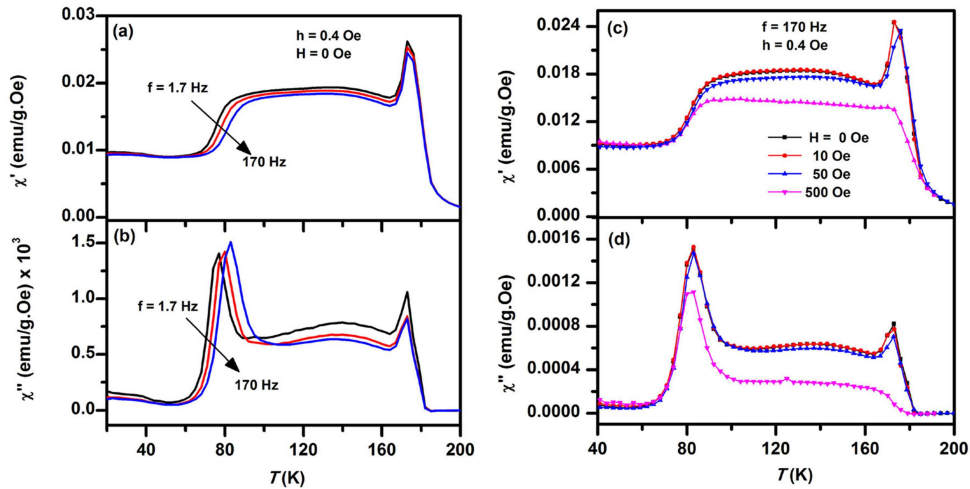}
\caption{\label{Fig.5 } (Color) Temperature dependence of the (upper panels) in-phase $\chi{'}$ and (lower panels) out-of-phase $\chi{''}$ components of the  ac-susceptibility recorded for (a,b) different frequencies $f$ = 1.7, 17, and 170 Hz; $h$ = 0.4 Oe, and (c,d) under different superimposed dc magnetic fields $H$ = 0, 10, 50 and 500 Oe; $f$ = 170 Hz;  $h$ = 0.4 Oe.}
\end{figure*}
\par In the susceptibility curves presented in the left panels [(a) and (b)] of Figure 5, the onset of the high temperature peak is frequency independent, as expected for a ferromagnetic transition, while the lower temperature peak is quite frequency dependent, suggesting a low-temperature glassy behavior. It is remarkable that, as observed in systems with blocked magnetic clusters, the magnitude of $\chi{''}$ ($T$) drops very rapidly below the low temperature peak and remains at and near zero, with negligible frequency dependence, on further lowering of the temperature. This behavior suggests that the fundamental entities of this glassy transition are groups of coherent spins (superspins) rather than individual atomic spins. \cite{Jonason19966507, Hansen20024901} The low temperature ac-susceptibility of spin glasses is weakly affected by low superimposed dc magnetic fields. This is also the case in LCMO18, as seen in the right panels [(c) and (d)] of Figure 5. A magnetic field of 50 Oe minorly affects the susceptibility curve in the whole measured range of temperatures. However, a field of 500 Oe significantly reduces the ac-susceptibility above 80 K. The ac-susceptibility of the low-temperature phase is essentially not affected by the superimposed magnetic field, akin to re-entrant ferromagnets.\cite{Jonason19966507}
\par Although disordered, spin glass materials undergo a magnetic phase transition at a temperature $T_g$ with well-established critical exponents. If a spin glass is cooled down below $T_g$, it will always be out of equilibrium, and its spin configuration will rearrange itself toward the equilibrium configuration for that temperature. \cite{Binder1986801,Jonsson2004174402} This equilibration is referred to as aging.\cite{Jonsson2004174402,Lundgren1983911,Kundu20064809} Aging can be observed in relaxation experiments, which can employ dc or ac excitation to record the isothermal magnetization or susceptibility as a function of time. \cite{Jonsson2004174402,Mathieu2002012411} If the temperature is changed, the system will again rearrange itself toward the equilibrium configuration of the new temperature; the system will be reinitialized or rejuvenated. Yet, it can be shown that the spin configuration resulting from the first equilibration is kept in memory while the second proceeds. \cite{Jonsson2004174402,Mathieu201067003} If the new temperature is lower than the initial one, the system equilibrates to configuration corresponding to new temperature and keeps the equilibrated configuration of the higher temperature in memory. On the other hand, if the new temperature is higher than the initial one, the system will be reinitialized, even for short durations at the new temperature. \par Some systems will first order magnetically as long-ranged ferro- or antiferromagnets, to become (or re-enter) a disordered phase at low temperatures. This is the case of re-entrant spin glasses, or in the case of ferromagnets, re-entrant ferromagnets.\cite{Mathieu2000441} The disordered phase is a consequence of the magnetic frustration in the ordered phase, hindering the perfect ordering. It has been shown that the low-temperature spin glass phase of such materials has similar dynamical behavior as those of ordinary spin glasses, and also that the ferromagnetic phase exhibits glassy features and aging, although the spin configuration of this phase reinitializes upon both positive and negative temperature cycling, unlike in spin glasses.\cite{Jonason199923} While the critical slowing down at the spin glass phase transition can be investigated by e.g. scaling analyses of the onset of $\chi{''}$ in ac-susceptibility measurements, \cite{Binder1986801} such analyses are difficult in re-entrant ferromagnets due to the ``parasitic" ferromagnetic phase contributing to the susceptibility. \cite{Jonason19966507,Mathieu2000441} Nevertheless, the onset of low temperature glass phase is indicated by the frequency dependence of the low temperature peak of the $\chi$-$T$ curves.

\par Aging phenomena are observed in LCMO18 at temperatures below 170 K. As illustrated in Figure 6(a), $\chi{''}$ relaxes downwards at constant temperature after being cooled from a reference temperature in the paramagnetic region ($T_{ref}$ = 220 K). Yet, the shape and magnitude of the relaxation of the $\chi{''}$($t$) curves are quite different at temperatures above and below the low temperature peak. The inset in Figure 6(a) shows the relaxation at 60 K in an expanded scale to clearly illustrate the difference in shape of the relaxation in the ferro and glass phases of the sample. The time-dependent susceptibility data recorded at various temperatures from 50 to 180 K is plotted (vertical lines) in Figure 6(b) and (c) as function of temperature. For comparison, a conventional temperature dependent ac-susceptibility measurement recorded using the same frequency and excitation ($f$ = 1.7 Hz, $h$ = 0.4 Oe) is also plotted. We can see in those panels that the weak downward relaxation observed at 50, 60 and 70 K start from about the same susceptibility values as those obtained in the ordinary temperature-dependent measurement. This nature of the aging related relaxation is similar to what is expected and observed to occur in spin glass states.\cite{Jonsson2004174402,Mathieu2002012411}
\begin{figure}
\includegraphics{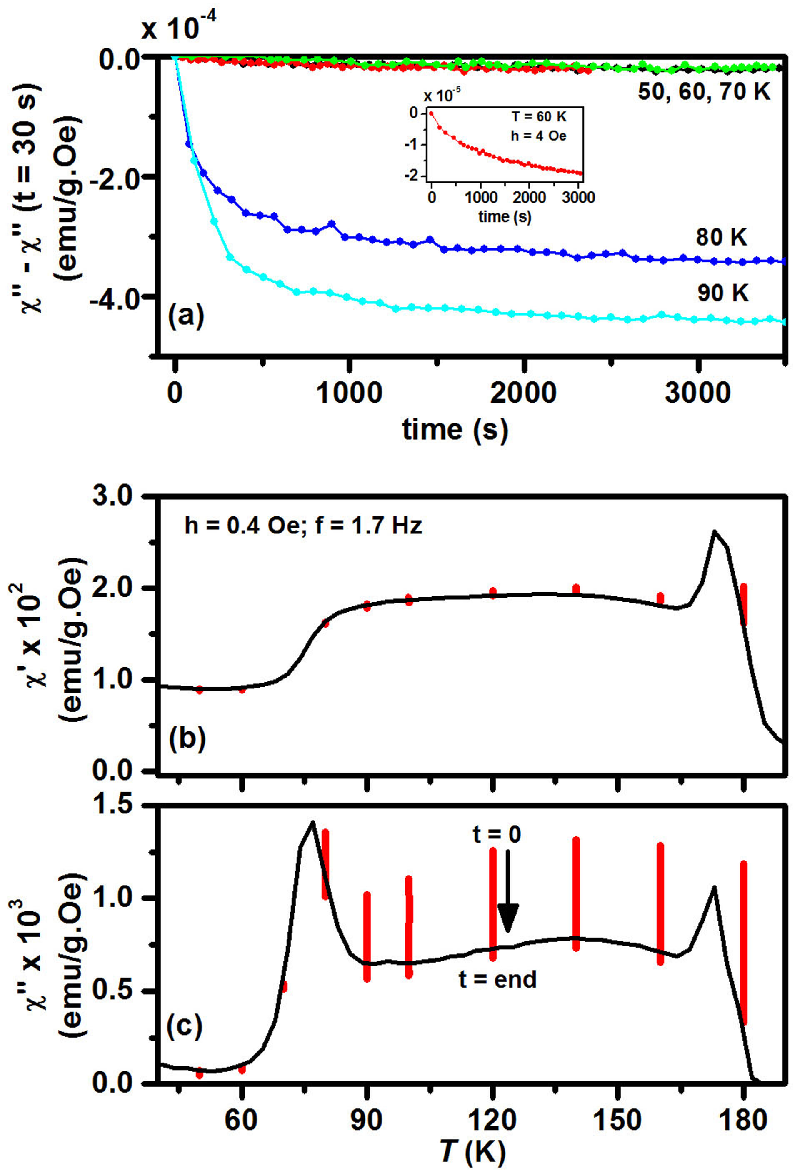}
\caption{\label{Fig.6 } (Color) (a): time-dependence of the normalized out-of-phase susceptibility for different temperatures recorded, after a quench from 220 K, at $T$ = 50, 60, 70, 80, and 90 K. Inset shows a similar curve measured at 60 K but using a higher field amplitude, $h$ = 4 Oe. (b) and (c): the time-dependent data presented in the upper panel is plotted as function of temperature, for the in-phase and out-of phase components of the susceptibility, without normalization. The top of the vertical lines corresponds to susceptibility value at $t$ = 0 s and the bottom corresponds to the end time (3600 s). Additional data obtained for $T$ = 100, 120, 140, 160, and 180 K are included. The results of an ordinary temperature- dependent measurement of the ac-susceptibility is included ($\chi{'}$($T$) and $\chi{''}$($T$) curves respectively ($f$ = 170 Hz; $h$ = 0.4 Oe).}
\end{figure}
For the temperatures above 70K, the behavior is quite different, and one can see (more clearly in fig. 6 (c)) that the relaxation starts from higher values of the ac-susceptibility, and finishes below the T-dependent values on the longest time observed in our experiments (3600 s). This behavior is reminiscent of that of the ferromagnetic phase of a re-entrant ferromagnet. \cite{Jonason199923} The difference in relaxation behavior between the two temperature regions (above and below 80 K) is again an evidence of the phase change from a frustrated ferromagnetically dominated response to a glassy one. \par Coming back to the origin of the magnetic behavior in these compounds, there are some plausible explanations\cite{Shenoy2007097201,Pai2003696,Ramakrishnan2004157203} such as the two electron fluid $lb$ model; one type of electrons are essentially localized combined with a distortion of MnO$_6$ octahedra (polaronic) while the other type are characterised by finite hopping and non-distorted lattice. One of the experimental proofs for such a scenario comes from the extended x-ray absorption fine structure (EXAFS) analysis.\cite{Booth1998853,Jiang2007224428,Bridges2010184401} The microscopic crystal structure is probed using EXAFS in several doped manganite samples/crystals which focuses on the fraction of Jahn-Teller (JT) distorted MnO$_6$ octahedra in the lattice; Mn$^{3+}$ is JT active while the Mn$^{4+}$ is JT inactive and in doped manganites there is a mixture of both. It is concluded that in the FMM samples the JT distortion is completely removed in the fully magnetized state of the sample which is achieved either by lowering the temperature or by application of external magnetic field. However, in the case of FMI samples there exist JT distorted MnO$_6$ octahedra even in the ferromagnetic state, nevertheless the application of relatively large magnetic field leads to reduction in the fraction of JT distorted MnO$_6$ octahedra.\cite{Jiang2007224428} These observations are linked to the fact that in FMM samples the $e_g$ electrons are completely itinerant and hence no Mn ions are JT active whereas in the FMI samples the $e_g$ electrons are localized between neighbouring Mn ions and hence some Mn ions are JT active. Due to such a combination of polaronic and non-polaronic MnO$_6$ octahedra in the FMI compositions the magnetic properties are also expected to be different from the ferromagnetic ground states that are observed in metallic compositions. The solid state nuclear magnetic resonance (NMR) experiments on LCMO samples have also pointed to the presence of different fractions of FMI and FMM phases depending on the composition and temperature. \cite{Papavassiliou2000761} Our results suggest that the magnetic and electrical properties with hole doping in low bandwidth manganites evolve from an antiferromagnetic insulator (AFI) at lowest doping levels via a frustrated ferromagnetic insulator (FFMI) for intermediate doping to the ferromagnetic metal (FMM) at higher doping.

\section{Conclusions}
We have reported the results of dc and ac magnetic characterization of a single crystal of La$_{0.82}$Ca$_{0.18}$MnO$_3$, which falls in the fundamentally interesting composition range of a ferromagnetic insulating phase in the family of manganites. The dc and ac magnetic measurements combinedly suggest that the low temperature magnetic state of this composition is a re-entrant ferromagnet. I.e. LCMO18 enters a frustrated but all ferromagnetic state at 170 K followed by a re-entrant superspin glass state at about 80 K that is transformed to a fully magnetized ferromagnetic state by a large enough magnetic field. This intrinsic glassy magnetic state is in accordance with the results of theoretical models and EXAFS experiments on the low doped manganites, which predict magnetic polaron formation. A correlated group of such polarons which make up the superspins can be termed as a nanoscopic phase separation.
\begin{acknowledgments}
Authors thank the Department of Science and Technology, Government of India and Swedish Foundation for International Cooperation in Research and Higher Education (STINT) for supporting this research. PAK, RM and PN thank the Swedish Research Council (VR) and the G\"{o}ran Gustafsson Foundation, Sweden for funding.
\end{acknowledgments}

%

\end{document}